%% file: paper.tex
\documentclass{Interspeech2024}

\usepackage{multicol}
\usepackage{multirow}
\usepackage{enumitem}
\usepackage{url,hyperref,cleveref}
\usepackage{diagbox}
\usepackage{booktabs}
\usepackage{pifont}

\renewcommand{\thefootnote}{}

\interspeechcameraready 

\title{Text-aware Speech Separation for Multi-talker Keyword Spotting}

\name[affiliation={1}]{Haoyu}{Li}
\name[affiliation={1}]{Baochen}{Yang}
\name[affiliation={1}]{Yu}{Xi}
\name[affiliation={1}]{Linfeng}{Yu}
\name[affiliation={1}]{Tian}{Tan}
\name[affiliation={2}]{Hao}{Li}
\name[affiliation={1}]{$^{\dagger}$Kai}{Yu}

\address{
  $^1$MoE Key Lab of Artificial Intelligence, AI Institute, X-LANCE Lab, Shanghai Jiao Tong University, China\\
  $^2$AISpeech Ltd, China}
\email{\{haoyu.li,baochen1202,yuxi.cs,ylf2017,tantian,kai.yu\}@sjtu.edu.cn, hao.li@aispeech.com}

\begin{document}

\maketitle

\renewcommand{\thefootnote}{}
\footnotetext{$^{\dagger}$Kai Yu is the corresponding author.}

\input{content/0_abstract}
\input{content/1_introduction}

\input{content/2_text-aware_multi-talker_kws}
\input{content/3_experimental_setups}
\input{content/4_results_and_analysis}
\input{content/5_conclusion}

\input{content/6_acknowledgements}

\bibliographystyle{IEEEtran}
\bibliography{refs}

\end{document}

%% file: content/0_abstract.tex
\keywords{multi-talker keyword spotting, text-aware speech separation, robustness}

\begin{abstract}

For noisy environments, ensuring the robustness of keyword spotting (KWS) systems is essential. While much research has focused on noisy KWS, less attention has been paid to multi-talker mixed speech scenarios. Unlike the usual cocktail party problem where multi-talker speech is separated using speaker clues, the key challenge here is to extract the target speech for KWS based on text clues. To address it, this paper proposes a novel {\em Text-aware Permutation Determinization Training} method for multi-talker KWS with a clue-based {\em Speech Separation} front-end (TPDT-SS). Our research highlights the critical role of SS front-ends and shows that incorporating keyword-specific clues into these models can greatly enhance the effectiveness. TPDT-SS shows remarkable success in addressing permutation problems in mixed keyword speech, thereby greatly boosting the performance of the backend. Additionally, fine-tuning our system on unseen mixed speech results in further performance improvement.


\end{abstract}

%% file: content/1_introduction.tex
\vspace{-5pt}
\section{Introduction}

Keyword Spotting (KWS), also known as Wake Word Detection (WWD), serves as a critical interface for enabling intelligent interaction on a vast array of edge devices. Despite the substantial advancements that have led to impressive performance benchmarks, KWS systems face significant hurdles in complex acoustic environments. These environments are often marred by environmental noise or overlapping interference from multiple speakers, posing substantial challenges to the system's efficacy. Such conditions significantly compromise the reliability of the wake-up functionality, underscoring the need for solutions to improve system robustness. A widely adopted strategy for improving robustness involves integrating a Speech Enhancement (SE) front-end before the KWS module~\cite{joint_se_wkp_2021,joint_vad_se_wkp_2023,noise_kws_Convmixer_icassp2022,joint_se_wkp_wav2vec_2022}. These works adopt joint training, curriculum training strategies, self-supervised wav2vec~\cite{wav2vec1.0}, etc., to enhance the resilience of KWS systems in noisy scenarios.



However, addressing environmental noise alone is insufficient for robust KWS systems. Interference from other speakers, i.e., overlapping speech, substantially impacts KWS performance more than ambient noise. Overlapping speech contains segments similar to the wake-up word and is challenging to eliminate without specialized design, which increases the risk of false alarms. On the other hand, if the false alarm rate is reduced by raising the threshold, the wake-up rate is greatly affected. This paper focuses on the robust KWS against multi-talker interference. Unlike the usual cocktail party problem where multi-talker speech is separated according to speaker clue, the focus here is to extract the correct speech channel containing the target keyword. To the best of our knowledge, little prior work has explicitly addressed the KWS task-specific multi-talker problem. 

\begin{figure*}[t]
    \centering
    \includegraphics[width=0.9\textwidth]{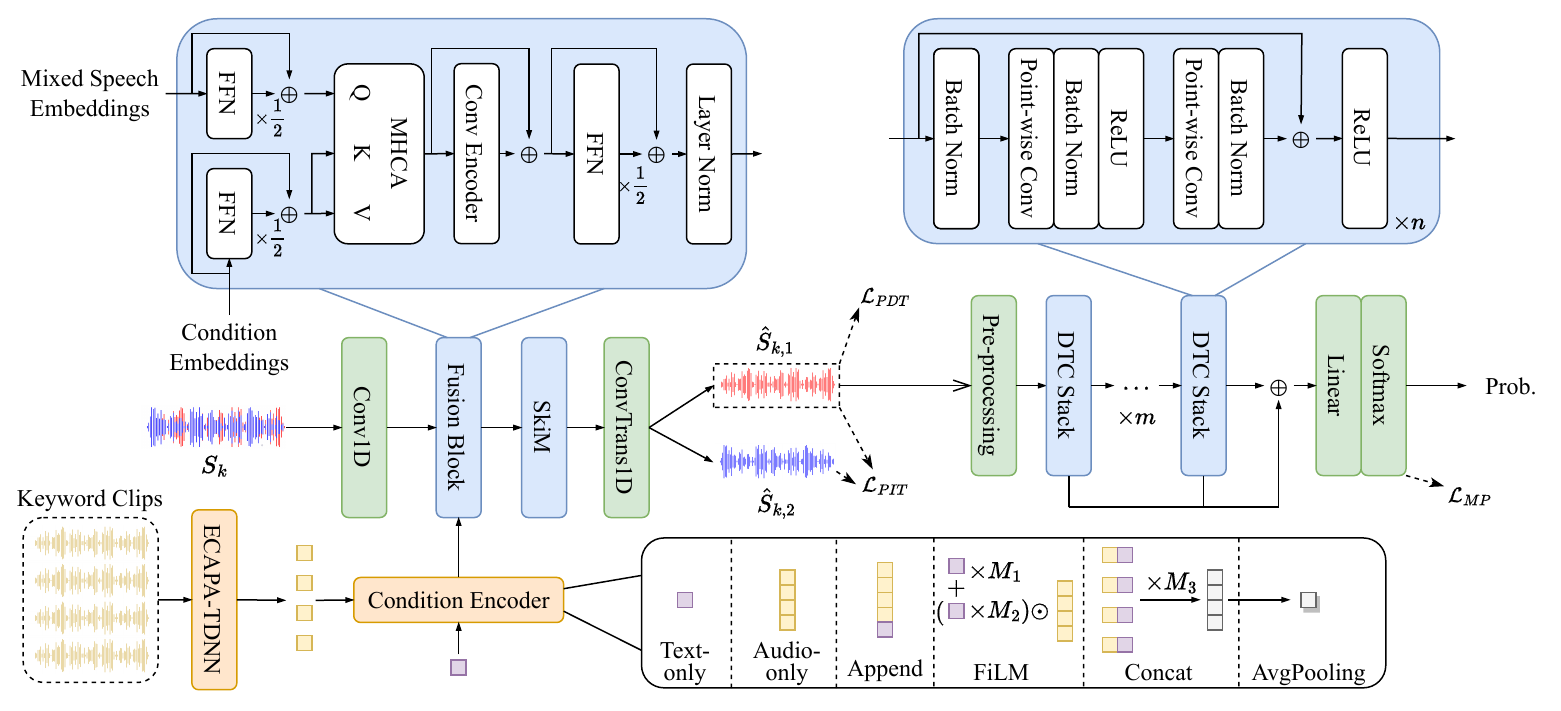}
    \caption{The overview of the whole system: TPDT-SS front-end(TSkiM) and MDTC KWS backend.}
    \vspace{-10pt}
    \label{fig:tskim}
\end{figure*}

In~\cite{text_dependent_SE_for_KWS}, the authors introduce TDSE, a SE front-end designed to simultaneously eliminate ambient and human noise for KWS. Each training sample for TDSE is synthesized with a keyword to ensure targeted noise suppression and enhanced keyword recognition. However, the absence of negative training data makes the model susceptible to overfitting, resulting in a high false alarm rate. As for multi-talker overlapping speech, training a speech separation model using PIT~\cite{pit_signal_1, pit_signal_2} criterion is a highly effective technique. It calculates the loss of all possible permutations of the separated signals and selects the permutation with the minimum loss to optimize the model. PIT is initially proposed for signal-level separation and subsequently applied to multi-talker automatic speech recognition (ASR)~\cite{pit_asr_1,pit_asr_2,pit_mimo_1,pit_mimo_2,pit_mimo_3}. While PIT has demonstrated superior separation performance, directly applying it for multi-talker KWS is unreasonable. The main reason is PIT regards all output channels equally, but for the KWS task, we only care about the keyword output channel. Recently, keyword clue-based frameworks~\cite{xiyu-clue-based-1, clue-based-2, clue-based-3,xiyu-clue-based-4,clue-based-5,dccrn-kws-2023} are widely explored under different sub-tasks of KWS, like open-vocabulary KWS or noise robust KWS. The keyword clue information can effectively bias the output of models and boost performance.  If we feed separated speech from all channels to the KWS model instead of the biasing channel, the processed data of the backend will double (assuming the number of output channels is 2). This increases not only the risk of more false alarms but also the computational burden of the backend, which is unsuitable for an on-device system. Thus, we propose the TPDT-SS front-end, a text-aware speech separation model tailored for noisy multi-talker KWS. Our contributions are summarized as follows.

\def\thefootnote{1}\footnotetext{\url{https://github.com/GnafiY/TPDT-SS-KWS}}
\begin{itemize}
    \item We demonstrate the importance of a speech separation front-end for multi-talker KWS and propose a novel TPDT-SS approach to enhance SS by incorporating keyword clues and permutation determinization training (PDT). Furthermore, we explore and identify potential methods for integrating keyword information into SS. Codes are open-sourced here$^1$.
    \item We demonstrate that channel permutation, which significantly impacts the KWS performance, is effectively mitigated by PDT. We observe a significant improvement in KWS tasks on audios processed by TPDT-SS.
    \item We fine-tune the KWS model using unseen multi-talker speech processed by TPDT-SS, further improving KWS performance. This demonstrates the potential of fine-tuning KWS models through the real mixed data from our proposed front-end.
\end{itemize}


%% file: content/2_text-aware_multi-talker_kws.tex
\vspace{-5pt}
\section{Text-aware multi-talker KWS}

 A text-aware multi-speaker system includes a TPDT-SS front-end and a KWS module. Using the keyword as a clue, the TPDT-SS front-end isolates the clear keyword speech into the preset channel while distributing the remaining speech to another channel under the assumption of two speakers. If the multi-speaker speech lacks the keyword, traditional PIT-based speech separation techniques are applied. The KWS module then processes the clear speech from the preset channel to determine the presence of the keyword. Further information is available in~\Cref{fig:tskim}.

\vspace{-5pt}
\subsection{The overview of TPDT-SS}
\vspace{-3pt}
\label{sec:overview_tpdt_ss}

From an architectural standpoint, the proposed system's primary distinction lies in adding a keyword condition module, which extracts the desired keyword speech from mixed speech. This module's design draws inspiration from incorporating target speaker embeddings into mixed speech for target speaker extraction (TSE) tasks, as outlined in~\cite{tse_overview}. Specifically, we introduce a keyword condition encoder designed to encode various types of keyword modality information, such as text-based wake words or a specific number of wake word audio clips. This keyword information is then integrated with mixed speech embeddings through an attention-based mechanism, which is essential for embedding the keyword clue into the SS process. Subsequently, the front-end module proceeds as a standard SS model, segregating the audio of different speakers into separate channels. For processing mixed speech, we employ SkiM~\cite{skim}, a model recognized for its lightweight design, low latency, and real-time SS capabilities, making it well-suited for KWS.

\vspace{-5pt}
\subsection{Keyword information extraction}
\vspace{-3pt}

We integrate diverse numbers, modalities, and merging functions within the front-end to maximize the efficacy of the keyword condition module. We create an embedding vector to represent textual keyword clues, and for audio keyword clues, we employ the ECAPA-TDNN module~\cite{ecapatdnn}. Both the textual and audio clue vectors are trained from scratch alongside the other components of the TPDT-SS system. In the subsequent sections, which focus on experiments and results, our front-end model, TPDT-SS, integrates the SkiM module, referred to as Text-aware SkiM (TSkiM). Our experiments leverage different types of clues, including text-only, audio-only, as well as combinations of text and audio clues. More details are provided in the bottom portion of Figure~\ref{fig:tskim}.

\vspace{-5pt}
\subsection{MDTC KWS module}
\vspace{-3pt}

We utilize the Multi-scale Depthwise Temporal Convolution (MDTC)\cite{mdtc} KWS model, designed to extract multi-scale features through varied receptive fields, as illustrated in the top right section of \Cref{fig:tskim}. To train the KWS backend, we employ a refined max-pooling loss~\cite{wekws}, a practical training criterion adapted from its initial formulation~\cite{maxpooling-loss}. Further specifics about this loss are detailed in~\cite{wekws}.

\vspace{-5pt}
\subsection{Training and inference}
\vspace{-3pt}
\label{sec:criteria}

The backbone of TPDT-SS can be divided into three parts: the text and/or audio condition encoder, attention-based conformer~\cite{conformer} fusion block, and SkiM~\cite{skim} separator as mentioned in~\Cref{sec:overview_tpdt_ss}. We adopt the time-domain SI-SNR loss\cite{timedomain-sisnr} to optimize the model. Specifically, for overlapping speech $S_{k}$, which is mixed by reference $S_{k,1}$ and $S_{k,2}$, we denote separated outputs by $\hat{S}_{k,1}$ and $\hat{S}_{k,2}$, where $\hat{S}_{k,1}$ always contains keyword if $S_k$ is a keyword-mixed speech and $\hat{S}_{k,2}$ is keyword-unrelated. 

Whether the training speech contains the keyword or not, the front-end model always applies the PIT criterion just like general SS models:
\vspace{-5pt}
\begin{equation}
\label{equ:pit}
\resizebox{0.91\hsize}{!}{
    $\mathcal{L}_{PIT}(S_k) =  \min_{\sigma}\{-\text{SI-SNR}(\hat{S}_{k,1}, S_{k,\sigma(1)}) - \text{SI-SNR}(\hat{S}_{k,2}, S_{k,\sigma(2)})\},$

}
\vspace{-5pt}
\end{equation}
where $\sigma$ denotes one of the two ($2!$) possible permutations. Then, the separation model is optimized by the permutation with minimum loss. For training data that contains the keyword, the model is additionally guided by $\mathcal{L}_{PDT}$ to separate the keyword speech to the preset channel and output residual parts to another channel. The scale-invariant signal-to-noise ratio (SI-SNR) loss of PDT can be formulated as follows:
\vspace{-3pt}
\begin{equation}
\vspace{-3pt}
\label{equ:pdt}
\resizebox{0.91\hsize}{!}{
    $\mathcal{L}_{PDT}(S_k) = -\text{SI-SNR}(\hat{S}_{k,1}, S_{k,1}) - \text{SI-SNR}(\hat{S}_{k,2}, S_{k,2}), $ 
}	
\end{equation}where the permutation of the output is preset. The total loss $\mathcal{L}$ of $S_k$ is the summation of the two kinds of training data:
\begin{equation}
\label{equ:final}
    \mathcal{L}(S_k) = \mathcal{L}_{PIT}(S_k) + y_k\mathcal{L}_{PDT}(S_k),
\end{equation}
where $y_k$ equals 1 if $S_k$ contains keyword, otherwise 0.

%% file: content/3_experimental_setups.tex
\vspace{-5pt}
\section{Experimental setups}


This section outlines the construction of datasets and the setup for both the TPDT-SS front-end and the MDTC KWS backend. Subsequently, we detail the KWS performance metrics employed for evaluating the multi-speaker KWS task. The baseline SkiM is available as an open-source tool in ESPnet-SE~\cite{espnet-se}.

\vspace{-5pt}
\subsection{Dataset}
\vspace{-3pt}
\label{dataset_description}
There are three parts of data to evaluate the performance: mixed multi-talker general ASR data, environmental noise, and mixed keyword data. The following are the details of the used datasets.

\begin{itemize}
    \item \textbf{Libri2Mix (L2M)}. Libri2Mix~\cite{librimix} is a speech separation corpus with ambient noise. It is derived from LibriSpeech~\cite{librispeech} clean subset and WHAM!~\cite{wham} noise. Each sample in Libri2Mix is synthesized
    by two different audios from LibriSpeech and a piece of ambient noise from WHAM!. We follow the official scripts to prepare general multi-talker ASR data.
    
    \item \textbf{Snips2Mix (S2M)}. Snips2mix is a self-constructed multi-talker KWS dataset containing the keyword ``Hey Snips'' in each sample. Hey Snips~\cite{snips} is an open-source KWS dataset that specifically uses ``Hey Snips'' as the keyword. Each sample in Snips2Mix is constructed by mixing one item from each of three datasets: LibriSpeech clean part, Hey Snips positive part, and WHAM! noise. We implement our simulated scripts based on the Libri2Mix data preparation scripts. They are open-sourced along with the codebase.

    \item \textbf{Snips2Mix-2000 (S2M-2000)}. This is an additional 2000 training pieces of training data simulated with the same scripts as Snips2Mix, designed to mimic the unseen data generated by users. This data has no clean audio reference and is used to fine-tune the backend to further validate the model's generalization and potential.

    \item \textbf{Hey Snips (Snips)}. We have mentioned the Hey Snips dataset in Snips2Mix before. We also use the original clean Hey Snips dataset to train our backend MDTC KWS model. The details of the dataset can be found in~\cite{snips}. Unless otherwise noted, our base KWS models are trained on the clean Hey Snips dataset.
\end{itemize}

To create a multi-speaker dataset for SS, we combine Libri2Mix and Snips2Mix. For the evaluation of KWS models, a test dataset is constructed, comprising 4.2 hours of negative samples and 1.9 hours of positive samples. Detailed information about this dataset is provided in~\Cref{tab:dataset}.

\begin{table}[b]
  \centering
    \vspace{-10pt}
    \caption{The number of utterances of datasets. ``-" means none and ``/" represents positive/negative.}
  \begin{resizebox}{1.0\columnwidth}{!}
  {

    \begin{tabular}{c c c  c c }
        \toprule
        \textbf{Dataset} & \textbf{Model} & \textbf{Train} & \textbf{Dev} & \textbf{Test}\\
        \midrule
        L2M & \multirow{2}*{TPDT-SS} & 13900 & 1500 & 3000 (4.2hrs)  \\
        S2M &  & 5000 & 1500 & 3000 (1.9hrs) \\
        \midrule
        Snips & \multirow{2}*{MDTC} & 5799/44859 & 2484/20179 & 2529/20543 \\
        S2M-2000 &  & 2000 & - & - \\
        \bottomrule
    \end{tabular}
    }
   \end{resizebox}
   \label{tab:dataset}
   \vspace{-10pt}
\end{table}

\vspace{-5pt}
\subsection{TPDT-SS front-end configurations}
\vspace{-3pt}
SkiM open-sourced in ESPnet-SE toolkit~\cite{espnet-se} serves as the baseline for SS. The TPDT-SS front-end is developed using the ESPnet toolkit~\cite{espnet}. Each SS model undergoes training for up to 100 epochs, employing a batch size of 64 and chunk iterators set to 24,000. The training process utilizes the Adam optimizer~\cite{adam} with a learning rate of 1e-3. The configurations for all hyper-parameters used during training, and all training details, along with the codes, are published in the codebase.

The condition list generated by randomly selecting audio segments from Hey Snips is fixed during training and inference inspired by the prior work~\cite{dccrn-kws-2023}. We conduct experiments with 0, 10, 20, 50, and 100 keyword clips as audio conditions to find the optimal condition option for TPDT-SS. Each TSkiM model in~\Cref{sec:results} is trained with text conditions. For example, TSkiM represents the text-only condition, and TSkiM-10 represents the combination of text and 10 fixed audio segments. By the way, the ECAPA-TDNN module is discarded after extracting speech conditions for inference, which saves a lot of memory and computational resources. 


\vspace{-5pt}
\subsection{MDTC KWS backend configurations}
\vspace{-3pt}
We leverage the MDTC KWS backend implemented in WeKws\cite{wekws} and use the default configuration. The default backend KWS model is trained on the Hey Snips dataset. We evaluate the backend performance through the metrics: recall (Rec) and false alarms per hour (FA/h). Recall comparison is conducted under 0.5 FA/h in the following results section. To better compare the KWS performance for different SS models, we employ two selection methods to merge backend results: 
\begin{itemize}
    \item \textbf{Max Selection (MAX)}. We apply the keyword spotting module to both output channels, and the maximum score will be selected as the final output. The cost is that the backend inference computation grows with the number of channels.
    \item \textbf{Channel-1 Selection (CH1)}. Since the TPDT-SS model has acquired the capability to determine the permutation, there is no need for extra channel selection. Thus, we will only apply the KWS module to the output of the first channel (preset as the keyword channel).
\end{itemize}

%% file: content/4_results_and_analysis.tex
\vspace{-5pt}
\section{Results and analysis}
\label{sec:results}

\subsection{The importance of SS front-end for multi-talker KWS}
\vspace{-3pt}

\begin{table}[b]
    \centering
    \vspace{-10pt}
    \caption{Results of KWS models trained on clean/mixed speech.}
    \resizebox{0.48\textwidth}{!}{
    \begin{tabular}{ c c c c}
        \toprule
        \multicolumn{1}{c}{\textbf{Training Data}} & \multicolumn{1}{c}{\textbf{Test Data}} & \textbf{Front-end} & \multicolumn{1}{c} {\textbf{Recall(\%)}} \\
        \midrule
        \multirow{2}{*}{Snips} & Snips & - & 97.43 \\
        & L2M + S2M  & -  & 19.23 \\
        \midrule
        \multirow{2}*{Snips + L2M + S2M} & \multirow{3}*{L2M + S2M} & -  &  82.83 \\
         &  & SE  & 83.63  \\ 
        Snips + L2M + S2M + S2M-2000 &  & -  & 76.10 \\
        \bottomrule
    \end{tabular}
    }
    \vspace{-10pt}
    \label{tab:kws_on_mix}
\end{table}

We first explore the feasibility of training the KWS backend on mixed speech. We can conclude from~\Cref{tab:kws_on_mix} that the KWS module trained on clean data performs exceptionally poorly on mixed speech, which indicates that the KWS model trained on clean data is not robust against noisy environment. Then, we train the KWS model on Hey Snips, Snips2Mix, and Libri2Mix datasets to test whether the KWS model converges on mixed speech data. Results notify us that the KWS module trained on mixed data also performs poorly on mixed data (82.83\% vs 97.43\%). We also adopt a naive speech enhancement (SE) version SkiM (only one output channel) trained by time-domain SI-SNR loss~\cite{timedomain-sisnr} like the work~\cite{dccrn-kws-2023}. Although the SE front-end works well in~\cite{dccrn-kws-2023}, adding speech noise in training data does not perform well (83.63\% vs 82.83\%). Besides, additional mixed data does not help; instead, it degrades the performance further by adding more noise (76.10\% vs 82.83\%). The results indicate the task is challenging to tackle with only the KWS backend or adding a SE front-end and requires an SS front-end to process multi-talker speech.

\begin{table}[t]
    \centering
     \caption{Results of TSkiMs on different audio-text information fusion methods. The number of audio clips on the front-end models is all set to 10.}
    \resizebox{0.48\textwidth}{!}{
    \begin{tabular}{c c  c c  c  c}
        \toprule
        \multirow{2}{*}{\textbf{Model}} & \multicolumn{2}{c}{\textbf{SS (S2M)}} & \multicolumn{2}{c}{\textbf{SS (L2M)}} & \multicolumn{1}{c}{\textbf{KWS}} \\
        \cmidrule(lr){2-3} \cmidrule(lr){4-5} \cmidrule(lr){6-6}
        &   STOI & SI-SNR & STOI & SI-SNR & CH1(\%) \\
        \midrule
        Text-only & 85.96 & 9.71 & 84.52 & 9.52 & \textbf{92.80} \\
        Audio-only & 86.02 & 9.87 & 84.73 & \textbf{9.65} & 89.63 \\
        \midrule 
        Append & 86.15 & 9.95 & 84.58 & 9.55 & 91.60 \\
        FiLM ~\cite{perez2018film} & 84.23 & 9.30 & 84.16 & 9.31 & 91.30 \\
        Concat & \textbf{86.29} & \textbf{9.95} & \textbf{84.86} & 9.63 & 91.73 \\
        Concat + AvgPooling & 86.22 & 9.90 & 84.76 & 9.53 & 90.60 \\
        \bottomrule
    \end{tabular}
    }
    \vspace{-10pt}
    \label{tab:fusion_methods}
\end{table}

\vspace{-6pt}
\subsection{Keyword clue fusion methods}
\vspace{-4pt}
As shown in~\Cref{fig:tskim}, the fusion block takes audio and/or text clues of the keyword as the condition. Appropriate methods to fuse keyword-related information from the text and audio modals are worth exploring. We primarily examine four fusion methods, including text-only and audio-only.

We analyze the results in~\Cref{tab:fusion_methods} from two aspects. First, the text-only system gets the best KWS results (92.80\%). We believe that this is because text information is the most relevant and purest condition for the KWS task compared with audios, which contains much extra acoustic information, like record environments, speakers, etc. Most models with audio conditions achieve better separation results than the text-only system. This is because the richer information contained in audios is helpful for separation. Though the model with text-only conditions achieves the best KWS recall, we still think the front-end performance dramatically impacts improving KWS performance. Therefore, we conduct further experiments on the model with the concatenation (concat) fusion method.
\vspace{-5pt}
\subsection{Performance of KWS with TPDT-SS}
\vspace{-5pt}

In this section, we evaluate the proposed TSkiM, i.e., KWS  with the proposed TPDT-SS,
on datasets Libri2Mix (L2M) and Snips2Mix (S2M). In the upper part of~\Cref{tab:separation}, we evaluate the importance of the loss derived in~\Cref{equ:final}. From the lines of model (A) and model (C), we see that although the results of the front-end trained by PIT-only are good, the KWS results are not excellent enough compared with model (D)-(H). Besides, it's worth noting that the recall of channel one is nearly half of the max selection method, which indicates the nature of PIT: regarding the two output channels equally. That's why we need PDT. We also train the model (B) with only PDT and without PIT. The output channel for keywords is fixed, as the results are nearly the same between MAX and CH1. However, the front-end performance degrades dramatically, especially on Libri2Mix (SI-SNR=-0.47). Therefore, PIT is also indispensable in improving the results of general data. The model (D), which is trained with PIT and PDT, has relatively acceptable front-end and backend performance. In the bottom part, we further investigate the optimal number of audio clues and find that around 50 is the best. Another reasonable finding is that backend performance is almost positively correlated with front-end performance: the higher the separation metrics, the better the performance of the KWS system.


\begin{table}[t]
    \centering
    \caption{Performance of SS models trained on both \textbf{S2M} and \textbf{L2M}. TSkiM-$n$ means the fusion block of TSkiM encodes $n$ keyword clips as audio conditions. All front-end models use the same KWS backend trained on clean Hey Snips.}
    \resizebox{0.48\textwidth}{!}{
    \begin{tabular}{c c c c  c  c c  c c c}
        \toprule
        \multirow{2}{*}{\textbf{ID}} & \multirow{2}{*}{\textbf{Model}} & \multirow{2}{*}{\bm{$\mathcal{L}_{PIT}$}} & \multirow{2}{*}{\bm{$\mathcal{L}_{PDT}$}} & \multicolumn{2}{c}{\textbf{SS (S2M)}} & \multicolumn{2}{c}{\textbf{SS (L2M)}} & \multicolumn{2}{c}{\textbf{KWS}} \\
        \cmidrule(lr){5-6} \cmidrule(lr){7-8} \cmidrule(lr){9-10}
        & &  & &   STOI & SI-SNR & STOI & SI-SNR & MAX(\%) & CH1(\%) \\
        \midrule
        (A) & SkiM\cite{skim} & \ding{51} &  N/A & \textbf{86.04} & \textbf{9.93} & \textbf{84.93} & \textbf{9.74} & 82.80 & 41.77 \\
        (B) & TSkiM & \ding{56} &  \ding{51} & 80.38 & 5.82 & 66.23 & -0.47 & 88.53 & 88.47 \\ 
        (C) & TSkiM & \ding{51} & \ding{56} & 85.91 & 9.78 & 84.62 & 9.59 &  79.70 & 40.13 \\
        (D) & TSkiM & \ding{51} & \ding{51} & 85.96 & 9.71 & 84.52 & 9.52 & \textbf{92.20} & \textbf{92.80} \\
        \midrule 
        (E) & TSkiM-10 & \ding{51} & \ding{51} & 86.29 & 9.95 & \textbf{84.86} & 9.63 & 92.07 & 91.73 \\
        (F) & TSkiM-20 & \ding{51} & \ding{51} & 86.21 & 9.76 & 84.62 & 9.52 & 90.83 & 92.70 \\
        (G) & TSkiM-50 & \ding{51} & \ding{51} & \textbf{86.33} & \textbf{10.00} & 84.76 & \textbf{9.66} & \textbf{94.90} & \textbf{95.27} \\
        (H) & TSkiM-100 & \ding{51} & \ding{51} & 86.15 & 9.91 & 84.76 & 9.59 & 90.23 & 90.87 \\
        \bottomrule
    \end{tabular}
    }
    \vspace{-10pt}
    \label{tab:separation}
\end{table}

\vspace{-5pt}
\subsection{Fine-tuning on seen and unseen data}
\vspace{-3pt}
    Since our default KWS model is trained on the clean dataset and inference with cleaned data, a mismatch exists, and the model could be further optimized. This section investigates how the backend model performs after fine-tuning on the separated data. To generate the fine-tuning dataset for the KWS model, we run inference on the corresponding SS model ((A), (G) front-ends in~\Cref{tab:separation}) with datasets S2M (seen) and S2M-2000 (unseen). Fine-tuned KWS models are all initialized from the clean checkpoint ((A), (G) backends in~\Cref{tab:separation}). We also add L2M data processed by SS models to the fine-tuning dataset to avoid overfitting the keyword.
    
\vspace{-5pt}
\begin{table}[htbp]
    \centering
    \caption{Recall(\%) of fine-tuning KWS models by separated seen and unseen data.}
    \resizebox{0.48\textwidth}{!}{
    \begin{tabular}{c  c   c  c}
        \toprule
        \multicolumn{1}{c}{\textbf{Model}} & \multicolumn{1}{c}{\textbf{No FT}(\%)} & \multicolumn{1}{c}{\textbf{FT on seen data}(\%)} & \multicolumn{1}{c}{\textbf{FT on unseen data}(\%)} \\
        \midrule
        SkiM  & 82.80 & 93.27 & 80.73 (CH1) or 93.67 (MAX) \\
        TSkiM-50  & \textbf{95.27} & \textbf{96.77} & \textbf{97.57}  \\
        \bottomrule
    \end{tabular}
    }
    \label{tab:kws_ft}
    \vspace{-10pt}
\end{table}

From~\Cref{tab:kws_ft}, we can see the consistent improvement of the models when fine-tuning on the seen training data inferred by the corresponding front-end models. However, with the addition of unseen data, models behave differently. The new training data generated from TSkiM are mostly keyword clips as the majority of keyword audios are separated into the preset channel. Thus, the system has been further significantly improved. Conversely, SkiM does not have this advantage, so we have to randomly select a channel (CH1) or utilize the KWS model to score each separated segment and choose the higher one manually (MAX). The two selection methods do not perform well. MAX selection has a slight gain but needs the backend system to score each segment, which adds the computational burden. The CH1 selection method is not a reasonable way to pick the data, so the fine-tuning result with the online data is even worse than the clean result. The results in~\Cref{tab:kws_ft} indicate the potential and generalization of our proposed model.

\textbf{Limitations.} The front-end SS model increases the computational resource and memory consumption, which is a burden for the device system. In the future, we aim to conduct joint training between front-end and backend models, making TPDT-SS better fit the KWS task. 

%% file: content/5_conclusion.tex
\vspace{-5pt}
\section{Conclusion}
\vspace{-3pt}
    In this paper, we propose TPDT-SS, a novel text-aware speech separation model for multi-talker KWS. First, we evaluate the importance of the SS module. Then, we adopt multiple ways to construct keyword clues to bias the output of the front-end SS. The proposed PDT is effective at fixing the output channel of keyword speech. Compared with the original SS baseline trained with only PIT, the proposed model not only saves the computational resource as it only infers one channel data for the KWS backend but also improves the SS and KWS performance significantly. In addition, our proposed model can further perform better by fine-tuning with actual online multi-talker KWS data, while cannot be done by PIT-based models, indicating the superior application prospect of our methods.

%% file: content/6_acknowledgements.tex
\section{Acknowledgements}

This work was supported by Shanghai Municipal Science and Technology Major Project (2021SHZDZX0102), and Key Research and Development Program of Jiangsu Province, China (Grant No. BE2022059).